# A Mixed User-Centered Approach to Enable Augmented Intelligence in Intelligent Tutoring Systems: The Case of MathAIde App

Guilherme Guerino, Luiz Rodrigues, Luana Bianchini, Mariana Alves, Marcelo Marinho, Thomaz Veloso, Valmir Macario, Diego Dermeval, Thales Vieira, Ig Bittencourt & Seiji Isotani

Taylor & Francis
Taylor & Francis Group

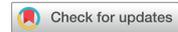 Check for updates

# A Mixed User-Centered Approach to Enable Augmented Intelligence in Intelligent Tutoring Systems: The Case of MathAIde App

Guilherme Guerino[a,b], Luiz Rodrigues[a], Luana Bianchini[a], Mariana Alves[a], Marcelo Marinho[c], Thomaz Veloso[a], Valmir Macario[c], Diego Dermeval[a], Thales Vieira[a], Ig Bittencourt[a,d] and Seiji Isotani[a,e]

[a]Center of Excellence in Social Technologies, Federal University of Alagoas, Maceió, Brazil; [b]State University of Paraná, Apucarana, Brazil; [c]Federal Rural University of Pernambuco, Recife, Brazil; [d]Harvard Graduate School of Education, Cambridge, USA; [e]Graduate School of Education, University of Pennsylvania, USA

**ABSTRACT**

This study explores the integration of Augmented Intelligence (AuI) in Intelligent Tutoring Systems (ITS) to address challenges in Artificial Intelligence in Education (AIED), including teacher involvement, AI reliability, and resource accessibility. We present MathAIde, an ITS that uses computer vision and AI to correct mathematics exercises from student work photos and provide feedback. The system was designed through a collaborative process involving brainstorming with teachers, high-fidelity prototyping, A/B testing, and a real-world case study. Findings emphasize the importance of a teacher-centered, user-driven approach, where AI suggests remediation alternatives while teachers retain decision-making. Results highlight efficiency, usability, and adoption potential in classroom contexts, particularly in resource-limited environments. The study contributes practical insights into designing ITSs that balance user needs and technological feasibility, while advancing AIED research by demonstrating the effectiveness of a mixed-methods, user-centered approach to implementing AuI in educational technologies.

## 1. Introduction

Artificial Intelligence in Education (AIED) aims to improve learning experiences by integrating AIED systems into the educational domain. The benefits of AIED include personalized learning experiences, increased student engagement, tailored feedback, and improved retention rates (Chen et al., 2022; Crompton et al., 2024). Additionally, AIED systems can scale up educational resources to reach a wider audience, promoting a more inclusive and accessible learning environment (Memarian & Doleck, 2023; Vinuesa et al., 2020) when well designed. AIED systems encompass a range of applications, such as Intelligent Tutoring Systems (ITS) and Adaptive Systems. The literature provides consistent evidence of the contributions of these systems to education (Hillmayr et al., 2020; Liu et al., 2020; Ma et al., 2014; Major et al., 2021). However, three main concerns must be addressed in AIED systems.

First, teachers are vital stakeholders of the AIED system. They decide which tools to use, when, and how to use them (Chiu et al., 2023). Therefore, involving teachers in the design process of AIED systems is essential to ensure that these systems align with educational goals, promote compelling learning experiences, address ethical concerns, and are of high usability (Dermeval et al., 2018a, 2018b; Topali et al., 2025). Unfortunately, there has been little AIED research from this perspective (Alfredo et al., 2024; Modén et al., 2021; Rodrigues et al., 2023). This research gap may lead to solutions that are not usable or fail to achieve educational goals.

Second, even AI tools such as GPT-4 have limitations despite their advanced capabilities. Large Language Models (LLMs) are known to suffer from hallucinations, which generate incorrect or nonsensical answers (Ji et al., 2023). Even well-established AIED research areas, such as Knowledge Tracing and Recommender Systems, have performance limitations (Abdelrahman et al., 2023; Rahayu et al., 2023). This limitation is crucial because teachers are unlikely to rely on AIED systems unless they trust them (Chiu et al., 2023).

Third, not everyone has access to the resources required to use the AIED systems, which demand technological resources such as computers and the Internet (Mousavinasab et al., 2021) as many regions lack access to such resources (Gasevic et al., 2018; Reimers, 2022). This issue led to research presenting an alternative interface for students to benefit from AIED systems, where teachers act as a proxy between the students and the system (Isotani et al., 2023). Instead of interacting with an ITS directly, through a keyboard or a touchscreen, students continue solving tasks on paper and sheets. Then, the teachers photograph the paper sheets using a mobile ITS, which assesses the student's solution and presents personalized feedback/recommendations for the teachers that, in turn, deliver those insights to the student (Davis et al., 2020; Patel et al., 2022; Silva et al., 2023). Notably, this alternative interface heavily depends on the ITS supporting handwritten instead of keyboard/touchscreen input.

However, a recent literature review revealed a substantial lack of ITSs for math education that supports handwritten input (Rodrigues, Pereira, et al., 2024). Implementing such an AIED system requires state-of-the-art AI features, which pose substantial technical challenges (Chiu et al., 2023). In response to this challenge, Augmented Intelligence (AuI) emerges as a viable solution (Broschert et al., 2019). AuI, also known as Intelligence Amplification, refers to the use of AI systems to enhance human intelligence rather than aiming for these systems to operate autonomously (Broschert et al., 2019). Recognizing the technical challenges of implementing advanced AI techniques within mobile ITS, AuI offers an alternative in which ITS suggests a solution to a given task (e.g., assessing a handwritten equation) but relies on human intelligence to provide a definitive answer and learn over time (Yau et al., 2021). For instance, a teacher could indicate that the ITS assessment was inaccurate and give the proper answer to it. The ITS can then adjust its suggestions and use this feedback to improve future analyses. However, to our knowledge, previous research has not explored AuI to enhance the design of mobile ITS supporting handwritten input (Alfredo et al., 2024; Topali et al., 2025).

Based on that context, the following research question (RQ) remains to be addressed:

- **How to design AuI so that it empowers mobile ITS supporting handwritten input?**

We address this RQ through a user-centered mixed methods approach involving designing, developing, and evaluating an AuI feature for MathAIde. Designed to support mathematics education, MathAIde is one of the few mobile ITS that supports handwritten inputs. By using multiple AI features – such as Object Recognition, Handwritten Math Equation Recognition, and Knowledge Tracing – MathAIde reviews mathematics exercises based on a photo, assesses the student's solution, and provides personalized feedback, learning analytics, and recommendations for teachers, which are in charge of forwarding these insights to students.

In addressing our RQ, we conducted four studies: a brainstorming session with 14 teachers, high-fidelity prototyping involving a team of UX experts, A/B testing with three teachers, and a case study in real classroom environments involving three teachers and 49 students. We considered all ethical precautions, which are described in the subsequent sections. As a result, we achieved an AuI design that is highly aligned with users' needs within the context of a real mobile ITS supporting handwritten input. Thus, we contribute to Human-computer Interaction by demonstrating how to employ user-centered mixed methods principles to improve the design of AIED systems concerned with being accessible to low-resource populations while attending to users' needs and handling technological challenges.

The remainder of this article is organized as follows: Section 2 shows related work as well as an explanation of the functionality of MathAIde; Section 3 provides an overview of the mixed method used in this research and ethical considerations; Section 4 presents the brainstorming procedure and results; Section 5 shows the procedure and results of high-fidelity prototyping; Section 6 presents the A/B testing procedure and results; Section 7 presents the procedure and results of the case study; Section 8 shows the discussion based on our findings, implications and threats to validity; Finally, Section 9 shows the conclusion.

## 2. Background

This section provides background information on AuI, then discusses related work and the app used to enable this research.

### 2.1. Augmented Intelligence

AuI is a subfield of AI that emphasizes enhancing, rather than replacing, human cognitive capabilities (Sadiku et al., 2021). The term originates from William Ross Ashby's Introduction to Cybernetics (1956), highlighting a fundamentally collaborative relationship between humans and machines. Unlike traditional AI systems, often designed to function autonomously, AuI systems aim to support and refine human decision-making processes. As a result, the distinction between artificial and augmented intelligence lies primarily in the decision-making authority (Broschert et al., 2019).

While conventional AI may operate independently, automating tasks such as spam filtering, language translation, or route optimization, AuI systems are inherently interactive and assistive. They synthesize data, detect patterns, and present options that empower humans to make more informed and precise choices. This partnership model has found particular utility in domains that require complex judgment informed by extensive data, such as medicine and finance, as found in the survey Yau et al. (2021). For example, the survey found applications in domains like biometrics (e.g., facial recognition), finance (e.g., managing risk and compliance), and healthcare (e.g., performing medical analyses).

On the other hand, the same survey found the educational domain to be a promising future application of AuI Yau et al. (2021). As a response, research has increasingly explored this line. For example, Toivonen et al. (2019) introduced an AuI-based method to facilitate the analysis and improve the performance of decision trees in educational data mining. Another application is introduced in Troussas et al. (2025), aiming to empower ITSs, where an AI system structures a learner's text input, processes it, and generates personalized feedback for the learner based on their specific needs. In contrast, a feedback evaluator assesses the quality of the generated input and determines its helpfulness to the learner. In that context, Troussas et al. (2025) discusses how AuI can amplify human potential when integrated with pedagogical principles, fostering more adaptive learning experiences by prioritizing human-machine collaboration. Similarly, Alvarez-Icaza and Huerta (2024) discusses how AuI can improve the personalization and efficiency of educational technology, given that it results in an interaction more tailored to one's behaviors and preferences.

In that context, human-in-the-loop approaches can be seen as a component within the broader framework of AuI. While AuI refers to the overall goal of enhancing and supporting human decision-making through AI Sadiku et al. (2021), human-in-the-loop represents one of the practical mechanisms by which this goal is achieved. By involving humans at challenging points in the AI system's workflow, such as during training, evaluation, or real-time decision-making (e.g., Rodrigues et al. (2023); Di Mitri et al. (2022); Dermeval et al. (2018a)), human-in-the-loop ensures that the system remains aligned with human judgment, values, and oversight. On the other hand, the human-in-the-loop approach is also used in contexts where augmentation is not the primary aim, such as in safety-critical monitoring or design processes. Therefore, while human-in-the-loop approaches can serve the goals of AuI, not all human-in-the-loop approaches are designed with augmentation or cognitive enhancement in mind, as in AuI systems.

### 2.2. Related work

This section discusses related work based on the literature review by Alfredo et al. (2024). We consider AIED studies related in that they i) include stakeholders in the design process and ii) explore AuI-related aspects in their investigated solution. Therefore, we assume Alfredo et al. (2024) as a suitable source for related work selection because it provides an overview of human-centered research in AIED and Learning Analytics and was published around a month before writing. Thus, it gives an up-to-date panorama of research aligned to concerns raised in Section 1.

Importantly, Alfredo et al. (2024) relies on the two-dimensional (i.e., user control and computational automation) Human-Centered Artificial Intelligence (HCAI)[1] framework Shneiderman (2022) to contextualize the level of human control and computer automation of AIED systems. In summary, they contextualize this framework to Learning Analytics/AIED systems in four quadrants (Q1, Q2, Q3, and Q4). Q1 represents systems with low user control and low computational automation, such as essential learning resources or report generators. Q2 concerns systems with high user control and low computational automation, where users can personalize information with manual operation, such as customizable visual dashboards or educator-driven analytics. Q3 depicts highly automated systems with low user control, such as predictive analytics or automated grading. Lastly, Q4 combines high user control with high computational automation, enabling manual operation and automated assistance, often seen in ITSs Alfredo et al. (2024).

Notably, we understand the HCAI framework as related yet different concepts. On the one hand, the HCAI framework provides a broader model to interpret how one might design AIED systems centered on human users. On the other hand, AuI establishes a particular way users might be included within AIED systems' functioning. Particularly, given the notion of AuI, we understand that AIED systems based on this approach enable a high level of user control, regardless of whether they offer low (Q2) or high (Q4) levels of computational automation.

Concerning Q2, Alfredo et al. (2024) found that the engagement of stakeholders, such as teachers and learners, is more passive than active in the design process. They are involved in the study (passive) but do not influence the design (active). Related to this study, research involving stakeholders in the design process engaged them in activities such as participatory design and exploratory tasks. Examples of systems within this quadrant are dashboards and learning design tools, which align with the limited computational automation (Alfredo et al., 2024). For instance, a relevant study in that line is Shreiner and Guzdial (2022).

Shreiner and Guzdial (2022) explored the design of a pedagogical support system to foster data literacy in social studies through participatory design sessions with both pre-service and in-service teachers. The system included instructional resources, standards-aligned lesson plans, and tailored data visualization tools contextualized for social studies. Through case studies with practicing teachers, the study identified key features that enhance technology tools' usability and perceived utility for integrating data visualizations into inquiry-based instruction. However, despite positive perceptions, actual adoption of the tools remained limited, revealing limitations in fully capturing the complexities of teacher adoption of these technologies in real-world educational contexts. Moreover, although such tools include educators in the design process and enable them to make decisions, as in AuI systems, they are not designed to explore advanced AIED techniques, such as automated recommendation and assessment.

Concerning Q4, Alfredo et al. (2024) found that stakeholders are more actively involved. Notably, the most frequent type of system in Q4 is ITS. While high computational automation is explained by ITSs' features (e.g., recommending learning activities or exercises), the level of human control likely concerns the stakeholders' ability to accept or not accept recommendations. One example is the study by Di Mitri et al. (2022), which presents CPR Tutor. CPR Tutor is a real-time multimodal feedback system that enhances cardiopulmonary resuscitation training by detecting performance errors using recurrent neural networks. It processes kinematic and electromyographic data to assess chest compressions against five key performance indicators. It delivers real-time audio feedback to address critical mistakes. Trained on expert data and validated through a field study with ten participants, the CPR Tutor demonstrates the effectiveness of multimodal, real-time feedback in improving cardiopulmonary resuscitation performance.

Another example is Lawrence et al. (2024), which explores how AIED systems can support teachers through shared control in orchestrating complex learning activities, particularly dynamic transitions between individual and collaborative learning. Through a secondary analysis of a co-design process involving middle school math teachers and participatory design methods, the researchers examined how educators conceptualize control, trust, responsibility, efficiency, and accuracy in working with an AI co-orchestration tool. The findings offer key insights into the human-AI interaction in classroom orchestration and demonstrate the value of human-centered learning analytics in informing the design of AI tools.

On the one hand, these systems exemplify AuI principles in advanced AIED systems, such as ITSs. As a result, they provide valuable contributions to the literature, such as advancing real-time tutoring systems by presenting a novel architecture and empirical evaluation of multimodal feedback in a training context (Di Mitri et al., 2022) as well as emphasizing teacher involvement in shaping AI integration and providing methodological reflections for eliciting meaningful feedback during the design of co-orchestrated educational technologies (Lawrence et al., 2024). On the other hand, previous research is limited to the design phase, failing to encompass implementation and evaluation or controlled experimental settings, which may not fully reflect the complexities and distractions of real-world learning environments (Di Mitri et al., 2022; Lawrence et al., 2024). Moreover, all those studies concern standard AIED systems, which might be hard to deploy in underserved regions, in contrast to those based on the alternative interfaces discussed before (Isotani et al., 2023).

To summarize, this section discussed research that explored user-centered design to create AIED systems with high user control in light of the literature review by Alfredo et al. (2024). Whereas some of those systems concern low computational automation, such as dashboards, a few studies have explored high levels of computational automation to design ITS-like systems that also offer high user control levels. Although these studies corroborate the idea of exploring user-centered design and AuI to create AIED systems, they do not approach the idea of alternative interfaces to make such advanced educational technology accessible to underserved educators. Therefore, to the best of our knowledge, no study has investigated how to employ user-centered design and AuI to design effective ITSs based on alternative interfaces such as handwritten input.

### 2.3. The MathAIde app

MathAIde is a mobile ITS supporting handwritten input focused on helping to teach mathematics Rodrigues, Guerino, et al. (2024). MathAIde's target audience is elementary school math teachers who are preparing their students' early math skills. The main flow of using MathAIde is as follows:

1. The teacher creates a list of exercises for a specific class. MathAIde has a base of ready-made math items.
2. The teacher asks students to solve the exercises on the list created in a paper-based.
3. In MathAIde, the teacher takes a photo of each student's notebook with the solved exercises.
4. Using computer vision and AI, MathAIde tells the teacher whether the student made a mistake or solved each exercise correctly.
5. MathAIde then recommends new exercises for the teacher to use based on student responses.

MathAIde incorporates the features of a recent paradigm named AIED unplugged Isotani et al. (2023); Silva et al. (2023). It encompasses using the Brazilian National Curriculum (BNCC) and adapting exercises for contextual relevance. It operates with minimum requirements to promote *conformity*. Regarding *disconnect*, MathAIde facilitates data collection and analysis without an internet connection and strategically uses online connectivity to update reports and improve AI models. A *proxy* component facilitates user interaction and technical processes, ensuring a user-friendly experience. For *multi-user* environments, the system recognizes the different roles of teachers, allowing detailed and personalized analyses for each student and supporting collaboration between them. Focusing on *unskillfulness*, the system prioritizes a user-friendly interface that is simple and abstract, reducing technical complications for educators and promoting accessibility for users with varying technological skills.

We investigated item 4) of the flow shown above for this research. The motivation for this focus was that computer vision and AI could make errors when detecting a student's answer (here called *misidentified student answers*), depending on the photo taken by the teacher, the student's writing, camera quality, and several other factors, given that handwriting math expression recognition remains an open problem in that context Aggarwal et al. (2023); Truong et al. (2024). For instance, at the time of writing, experimental results suggest the system's AI has an overall accuracy of around 70% in detecting all characters in a given answer and around 80% in detecting all but one character. Additionally, the AI system is limited to summation, subtraction, and multiplication equations with up to three digits Chevtchenko et al. (2023, 2024); Rosa et al. (2023). In this sense, we explored the concept of a mixed user-centered approach to enable AuI in MathAIde, allowing the end user (teacher) to work with the AI to overcome these errors that MathAIde may make due to technological restrictions and, thus, increase trustiness, reliability, and adoption Chiu et al. (2023).

## 3. Method

This research adopts a mixed user-centered approach to understand, prototype, and evaluate a functionality enabling AuI in MathAIde, an ITS that supports handwritten input as an alternative interface, enhancing accessibility in low-resource settings. Four distinct studies were conducted to achieve this goal. Figure 1 presents an overview of the mixed-methods approach.

The methodology unfolds through a series of studies addressing the challenge of correcting misidentified student answers by MathAIde. The first study was a Brainstorming session involving 14 participants who explored the problem and proposed solutions through group discussions. In the second study, Prototyping, a UX designer turned selected ideas into functional requirements and transformed them into interface prototypes. The third study, an A/B Test, involved three teachers performing tasks with the prototypes, with detailed logging and post-task interviews. Finally, the Case Study involved real classroom use by three teachers and 49 students, with all interactions logged and followed by interviews. Each study is detailed in its corresponding subsection.

### 3.1. Ethical considerations

All studies were approved by the University of São Paulo's Research Ethics Committee (CAAE 73285823.2.0000.5390). Specific ethical measures were adopted for each study:

- Study 1 – Brainstorming: we used the Informed Consent Form (ICF)[2] to inform participating teachers about ethical guarantees and their rights. All teachers read and accepted the term. This study's only collection was to characterize the participants and the ideas they generated in the brainstorming session. After collecting the data, the responsible researcher coded all the names of the 14 participants, sharing only the codes from P1 to P14 with the other researchers.
- Study 2 – Prototyping: As this is a study between members of the internal team involved in the project, we did not use the ICF, but rather a confidentiality agreement for participation in the project, read and signed by all members at the beginning. In this sense, the artifacts produced were disseminated among the members participating in this study; however, everyone knew the confidentiality involved in the project.
- Study 3 – A/B Test: In this study, we used ICF (same model as study 1), mentioning their rights and guaranteeing anonymity to the three participants. This study involved screen and voice recording. In this sense, after transcribing the audio into text, the responsible researcher coded the participants into P1, P2, and P3 so that he could analyze and share the data with the other researchers involved.
- Study 4 – Case study: All ethical precautions were taken in this study, which involved three teachers and 49 children. First, as it is a study in a real environment (in this case, schools), we collected approval[3] from the school management to make the environment available to carry out the studies. Second, we again applied and collected the ICF (same model as study 1) signatures from the three participating teachers. Third, as the 49 students were minors, we collected their parents' authorization so they could participate in the study. Data collection for this study was done using the photos taken with the application (step 3 from MathAIde's usage flow) without exploring any use of voice or image. After applying and collecting the data, everyone involved was coded by the responsible researcher: the teachers (P1 … P3) and the students (S1 … S49).

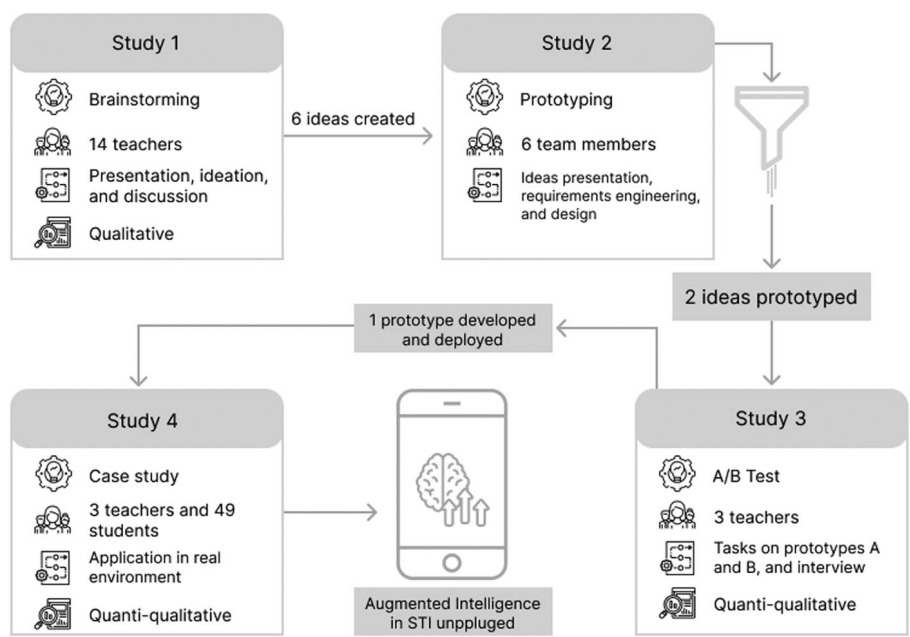

Figure 1. Overview of the mixed-methods approach.

## 4. Study 1: Brainstorming

In this first study, we used the brainstorming method with elementary school math teachers to explore how they could correct MathAIde's misdetection of a student's answer.

### 4.1. Participants

We recruited 14 teachers to participate in the study. Before starting any data collection, we explained the ethical issues of the research to the participants, and everyone agreed to the ICF. After signing the ICF, participants filled out a questionnaire about their demographic information, backgrounds as teachers, and use of applications in the classroom.

All participants were women and had a permanent contract at the school where they worked. The majority of participants were aged 55 or over ($n=9$), while the others were aged between 40 and 54 ($n=2$), 30 and 39 years old ($n=2$) or up to 24 years old ($n=1$). Most participants had 25 or more years of teaching experience ($n=8$), two were between 21 and 25 years, three were between 11 and 15 years, and one had 1 to 2 years of experience.

Regarding the use of digital applications in general, participants mentioned that they use basic applications such as messaging applications and social networks ($n=3$); basic and advanced applications such as banking applications and digital signatures, among others ($n=6$); basic and advanced applications and also applications that help with classroom work ($n=4$); and only one participant said that she does not use applications on her cell phone. Regarding the use of applications to support teaching practices, two participants mentioned that they had never used applications for this purpose; three participants mentioned that they had used them on specific occasions but at the moment they no longer use them; five participants currently use them in particular situations; and four use them frequently in classes. Some of the applications used by the participants to support teaching and school activities were an online class registration application, PARC Platform, Quizz, Google Drive, search tools, and YouTube.

### 4.2. Procedure

We divided Brainstorming into three stages, which will be explained in the subsections below.

#### 4.2.1. Stage 1. Presentation of MathAIde and the problem

After collecting information from the participants through the questionnaire, the brainstorming moderator presented the MathAIde application to the participants. Furthermore, the problem to be explored in Brainstorming was presented as: "*How to correct a student's answer that was misidentified by*

*MathAIde?*" We provided two examples of equations misidentified by MathAIde, illustrated in Figure 2, to support participants' understanding of the problem.

The two examples of student responses in Figure 2 show the misidentification of MathAIde. In example 1, the student solved the equation and wrote the answer correctly in the notebook; however, the application failed to identify the carry (also known as "carrying over" or overset), identifying the number 144 instead of 44 and mentioning that the student got the question wrong. In example 2, the student again did everything correctly in the notebook; however, the application did not correctly identify the cut made in the digit 5, identifying the number 1 instead of the cut number 5, also mentioning that the student got the question wrong. In both cases, the students performed the calculations correctly. However, the answers were classified as incorrect due to misidentification of the application's AI. In these cases, the teacher must intervene in the application to correct the data, providing greater user control and error correction.

### 4.2.2. Stage 2. Ideation

After characterizing the application and the problem, participants were challenged to propose innovative solutions to correct these misidentified equations. Participants were separated into three groups for a prior discussion of ideas and stimulation of creativity. The moderator provided sketches of cell phone screens, scissors, paper, pencils, erasers, and pens so participants could present their ideas. Each group could propose more than one idea.

### 4.2.3. Stage 3. Ideas presentation and discussion

After the three groups mentioned to the moderator that they had finalized their ideas' propositions, a representative from each group showed and explained the ideas to everyone. At this time, the moderator commented and answered questions about each proposal. The other participants commented and proposed modifications to the different groups' ideas.

## 4.3. Results

The entire brainstorming session lasted two hours. Six valid ideas were generated based on the participants' initial proposals, which were considered possible to integrate into MathAIde.

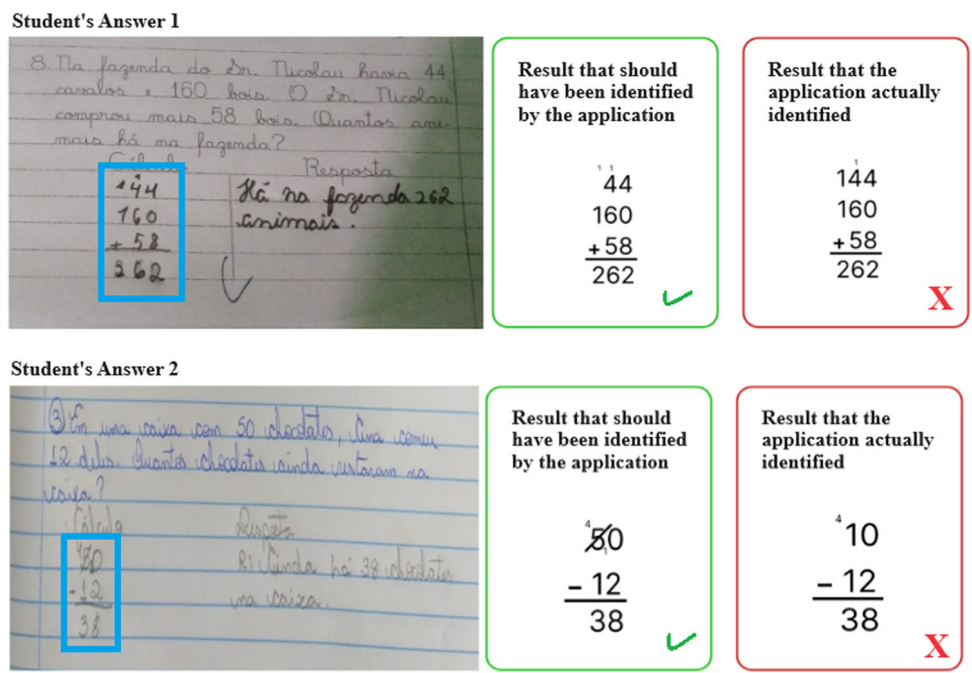

**Figure 2.** Examples used to support participants' understanding of the problem to be explored in brainstorming.

1. **Editing numbers.** This idea involves the possibility of editing numbers that MathAIde misidentified. In this sense, the participants suggested being able to edit a number by clicking on the wrong number and inserting the correct number. *Our analysis:* The idea presents a limitation when the error is not identifying a number but rather positioning that number in an equation, e.g., carrying over. One solution would be for the user to "tell" what that number is, e.g, when clicking on a number, the teacher would have the possibility to "Edit number" or "Resignify number" by clicking on the option that mentions "carry over."
2. **Fixed editing options.** This idea involves providing fixed editing options after a misidentification by MathAIde. After the user detects that the analysis is incorrect, MathAIde provides some pre-defined options, for example: "Carry over was not detected," "Cut number was not detected," and so on. When the user clicks on one of the options, MathAIde "re-analyzes" the equation with the bias of the teacher's selection made previously. *Our analysis:* For this idea, it is necessary to map the possible errors that MathAIde can make for different types of equations and train the AI to correct each type of error.
3. **Make a reservation before correction**. This idea involves the user mentioning some reservations before correcting, for example, "this equation must be done using carry over" or "this equation will have to cut numbers," among others. *Our analysis:* Making these reservations before all the equations may be unfeasible for learning purposes. However, these caveats may be identified for each exercise before corrections begin, e.g., in a particular exercise, the teacher knows that for the student to solve it correctly, it is necessary to use carryover. So, MathAIde already "expects" that students will use carry-over in the equation and tends to identify this characteristic. The idea is to use a "biased AI" depending on the recorded exercise.
4. **Have squares to identify numbers.** This idea involves delimiting the student's final equation with a previous structure formed by squares already on the answer sheet, where the student inserts each number within each square. *Our analysis:* This idea limits students' freedom to scribble, write, erase, etc. Despite being the "easiest" for MathAIde, it is unlikely to be accepted pedagogically.
5. **Multiple choice.** This idea involves asking multiple-choice questions with possible answers. In this sense, MathAIde would evaluate whether the choice is correct and not the equation. *Our analysis:* Getting the multiple-choice right does not mean the student got the calculation right. Furthermore, students in the initial literacy grades must learn about multiple-choice.
6. **Use colors to highlight numbers.** This idea involves using colors for MathAIde to communicate what each number represents to users. For example, numbers identified in the unit places would be blue, in the tens places, they would be green, the result would be yellow, and the carryover would be red. This way, the user could more easily identify where MathAIde is wrong and "change the color" of the wrong number. Changing the color would automatically change the function of the number. *Our analysis:* This idea makes it easier for the user to identify where MathAIde is going wrong using colors. The correction itself is very similar to idea 1, where the teacher can give new meaning to a number, changing, in this case, its color.

### *4.4. Synthesis*

The moderator collected all artifacts generated and sent them to the team members responsible for analyzing the findings. To choose the best options, the technical feasibility of developing the proposals and their alignment with the application's goal, both in terms of UX and the pedagogical issues involved, were taken into consideration.

## 5. Study 2: High-fidelity prototyping

This study aimed to analyze and prototype different solutions in high fidelity based on the best brainstorming findings.

### 5.1. Participants

The participants in this study were members of the internal team related to the project. Six people participated in this study: a UX designer, a requirements engineer, two UX researchers, and a UX lead. The UX designer was male, between 25 and 30 years old, with more than five years of experience in the area he works in. The requirements engineer was male, between 21 and 25 years old, and had worked for approximately one year in the requirements area. The two UX researchers were women between 30 and 40 years old, one with almost one year of experience in the area and the other with almost three years. The UX lead was male, between 25 and 30 years old, with 8+ years of experience in UX work.

### 5.2. Procedure

The study took place online through a virtual room on Google Meet. The UX lead started the meeting by presenting the objective of the meeting and the expected outcome. Then, the UX researchers presented the results achieved in the brainstorming study (Study 1) and listed the most viable solutions to be prototyped according to the objective of the application. After a discussion among all meeting participants, two options were selected based on the ideas and technical feasibility of development::

- Option A: Correction of numbers misidentified by AI.
- Option B: Correction of errors made by the student or indication that the student got the solution right.

The requirements engineer extracted the functionalities and business rules the next week. Therefore, another virtual meeting was scheduled, during which the requirements were presented and discussed with the other members. After the changes and refinement, the requirements were released for the UX Designer to prototype the interfaces and flows.

### 5.3. Results

Two prototypes were created in Figma to enable the correction of data misidentified by MathAIde. In prototype A (Option A), shown in Figure 3, MathAIde digitizes the identified calculation and shows it on the screen. The teacher can correct any number by interacting with the digitized buttons. Based on the teacher's exchange of numbers, MathAIde re-analyzes the calculation to provide a new result. In prototype B (Option B), also shown in Figure 3, the app shows the result obtained by the student (whether he got it right or wrong) and allows the following options: (i) if the student got it right but MathAIde shows that they made a mistake, the teacher can press the button saying that the student got it right and correct the AI result; (ii) if the student made a mistake, but MathAIde shows that he got it right, the teacher can trigger the errors that were made by the student and correct the AI result. In this sense, UX designer prototyped two options with different characteristics, one through digitalized calculation and the other through pre-defined options. Figure 3 only shows the MathAIde correction screen. However, to reach this screen, the teacher must follow an entire interaction flow, which includes taking a photo of the student's notebook, analyzing the result, and selecting to correct an error, if applicable.

### 5.4. Synthesis

The prototypes were presented to the other team members and validated. As there were no changes to be made, they were sent to carry out an A/B test.

## 6. Study 3: A /B test

With prototypes A and B finalized and validated, we carried out an A/B test to find the most appropriate solution according to the vision of our end users: teachers.

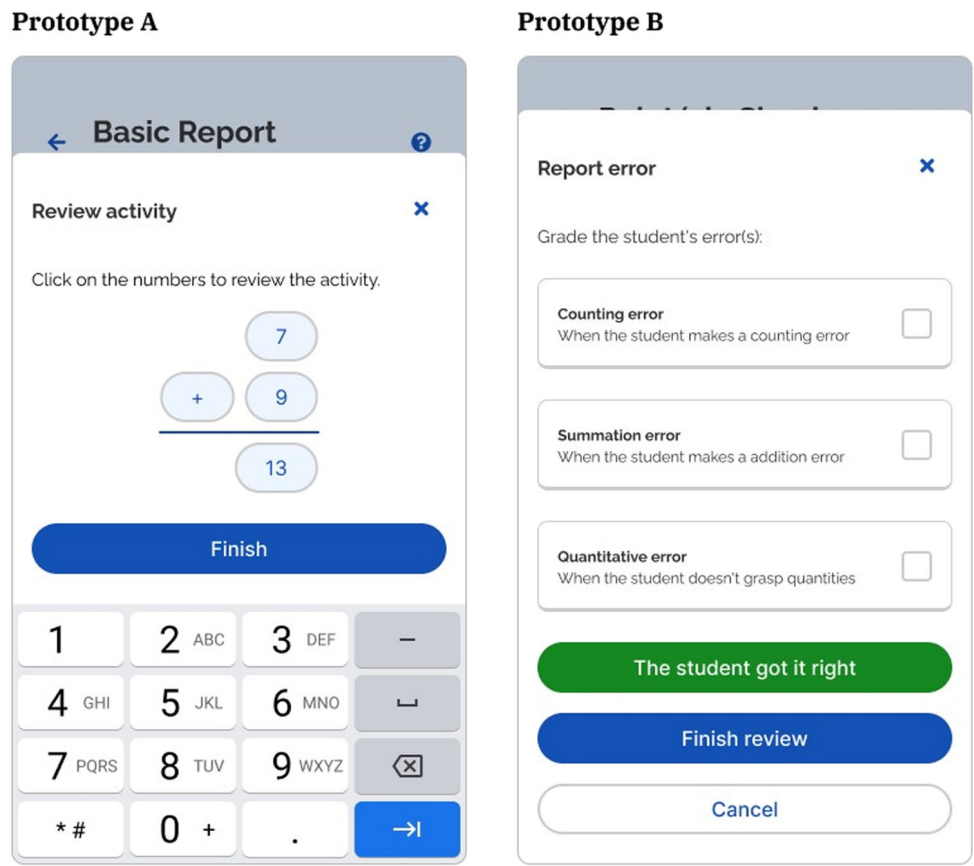

**Figure 3.** Screens of prototypes A and B.

## 6.1. Participants

We recruited three teachers to conduct A/B testing. None of these teachers participated in the brainstorming study. We explained all the rights and ethical guarantees to them, and they agreed with the ICF. In this study, two participants were men, and one was a woman. Two participants were between 25 and 29 years old, while one was between 50 and 54. Regarding classroom experience, all participants had between 6 and 10 years of experience. Concerning the use of applications in general, one participant said that he uses basic applications (e.g., messaging or social networking apps) and advanced applications (e.g., banking and digital signatures), while the other two mentioned that, in addition to basic and advanced applications, they also use applications that support their work. All participants mentioned that they use applications in teaching. Some applications mentioned are Canva, Gemini, Kahoot, and Super Games.

## 6.2. Procedure

Each participant met individually with two moderators in a virtual Google Meet room. After participants answered the ICF and the demographic questionnaire, the moderators explained the goal of the MathAIde and the objective of the test that would be carried out. After explaining and removing participants' doubts, the moderators sent the Figma link containing prototypes A and B. The exercise answer context to be recognized by MathAIde was as follows:

- **Question:** Larissa had 2 flowers and received 3 more from her mother. How many flowers did Larissa end up with?
- **Student answer:** $2 + 3 = 4$ (wrong answer)
- **AI identification:** $2 + 3 = 7$ (misidentification)

Based on this context, we asked participants to perform the following tasks:

- **Task 1.** In prototype A, record a student's answer and correct the number that MathAIde identified incorrectly (swap number 7 for number 4).
- **Task 2.** In prototype B, record a student's answer and report the error by selecting the type of error the student made (swap sum error for counting error).
- **Task 3.** In prototype B, record a student's answer and the report mentioning that the student got the question right (select that student got it right).

Before starting the tasks, we asked participants to share their screens. We then started recording to check the decisions made throughout the interaction and calculate the time taken for each task. In addition to the decisions made and the time spent on each task, after executing the tasks, we conducted a semi-structured interview to verify the participant's perception of each correction mode investigated. The questions are shown in Table 1.

### 6.3. Data analysis and results

To analyze the time in each task, we selected the exact moment in the recording when each participant started and ended the flow. We highlight that the flow for each task was complete; that is, it involved everything from registering the photo to selecting the option to correct the AI and correcting it. As for the interview questions, we transcribed the audio into text and analyzed each prototype, identifying speech patterns for each version used in the test.

Table 2 shows the results obtained regarding the time spent on each task. In terms of agility in task execution, the tasks carried out in prototype B had better results than the tasks carried out in prototype A. This result is understandable, given that prototype A has a greater need for user interaction, with a more significant number of clicks, greater possibilities for modifications, and, consequently, greater chances of errors. Prototype B, which has pre-defined options, does not require as much interaction but limits the teacher's possibilities for corrections. We also highlight a significant drop in the time spent on task 3 about task 2, on prototype B, which the learning bias can justify since participants may have learned to use the prototype in task 2, which facilitated the execution of task 3. We analyzed the participants' interview responses to understand each prototype more deeply.

#### 6.3.1. Prototype A

At first, P2 and P3 found it challenging to grasp the functionality of the buttons where the number $(2 + 3 = 7)$ appeared. They initially believed that when this screen appeared, they should simply proceed. However, upon realizing these buttons were clickable, P2 and P3 understood that they were meant to reflect the correct answer to the exercise $(2 + 3 = 5)$, not the student's original response $(2 + 3 = 4)$. P1, on the other hand, navigated through the feature without any issues.

The three participants liked this option because it was quick and efficient, and the student was not harmed (see citation C1). Among the positive points raised is an assessment that is more faithful to the student's response, and the AI can "learn" from these manual corrections (see citations C2 and C3). Of the negative points, the teacher will always need to finalize the answer (see citation C4). Participants also mentioned that they hope this does not happen so frequently and only stops with sporadic cases because it will take time if this adjustment is needed for all students (see citation C5).

Table 1. Questions used in the interview after performing the A/B test tasks.

| # | Question |
| --- | --- |
| 1 | What do you think of possibly adjusting the scanning done by AI? (Prototype A) |
| 2 | What are the positive points of prototype A? And the negatives? |
| 3 | What did you think of the possibility of correcting or pointing out an AI error from the "Report Error" menu? (Prototype B) |
| 4 | What are the positive points of prototype B? And the negatives? |
| 5 | Did any of the prototypes catch your attention positively? |
| 6 | Which of the prototypes did you feel was more efficient? |

Table 2. Time spent on each task.

|  | P1 | P2 | P3 | Average |
| --- | --- | --- | --- | --- |
| Task 1 | 1m40s | 2m33s | 2m34s | 2m15s |
| Task 2 | 45s | 52s | 1m05s | 54s |
| Task 3 | 1m10s | 34s | 20s | 41s |

- C1: "I thought it was very good, I believe this is a very efficient way for me to help the AI correct it. The student is not harmed." (P1)
- C2: "It improves the app, as the AI will understand the next times to recognize and for us for loyalty." (P3)
- C3: "It is an assessment that is more faithful to what the student knows, will have more precision in corrections for the next stages and groups, and will not harm the student." (P1)
- C4: "The teacher will always need to make a final review of the answers." (P1)
- C5: "It would take a lot of class time if I needed to make this correction for all my students." (P2)

### 6.3.2. Prototype B

P1 asked about the nomenclature "Report Error" because, according to him, this means that there is a "bug" in the application, and that is what it will report, and not report an error in the AI identification (see citation C6). P2 believes this option is more transparent about what should be done and makes reporting errors easier (see citation C7). The three participants also liked prototype B because it would not require much effort (since the reporting options are already pre-established) and would provide an accurate answer in the report that could help the evaluation (see citations C8 and C9). The main positive aspects raised are the agility in correcting and the teacher writing and directing the options (see citation C10). Of the negative points, it can take longer if there are many AI errors, and the term "Report Error" seems confusing.

- C6: "The term "report error" seems like I need to report a "bug" in the application to the developers." (P1)
- C7: "It made it easier because you would have an accurate answer in the report; as the errors appeared, it became clearer." (P2)
- C8: "I found it very valid. If I can correct it, for specific things, it won't require so much effort." (P1)
- C9: "I found it interesting because it provides feedback on how to improve the app and helps with our evaluation." (P3)
- C10: "The data collected would be more reliable, passing through 2 filters (AI and teacher). Because it is written, it directs the teacher." (P1)

### 6.4. Synthesis

Based on the results obtained for time spent and qualitative findings, we choose prototype B to promote AuI in MathAIde. In this sense, prototype B was coded and deployed in the MathAIde version to be used in the following study, which will be detailed in the next section.

## 7. Study 4: Case study in real environment

With the developed version of MathAIde enabling the correction of the AI represented in prototype B, we conducted a case study in a real environment to test the functionality.

### 7.1. Participants

We recruited three math teachers to use MathAIde in their classes. We emphasize that these three teachers also participated in the A/B Test, so they already had prior knowledge of the application. Teachers used the application with their 4th and 5th years classes, totaling 49 students. As mentioned in the ethical considerations section, in addition to the ICF applied to teachers, we collected consent from the schools where the two teachers work and the parents of each student who participated in this study and ensured that all ethical aspects were considered.

### 7.2. Procedure

We made MathAIde available for teachers to perform the study tasks. Each teacher used the MathAIde four times in their classes over four days. Each use of the MathAIde involved applying lists of exercises to students, recording each student's answers, and correcting the MathAIde if the app non-detected an answer. Figure 4 shows an example of the photo registration flow, from capture to possible correction. We emphasize that it would not always be necessary for teachers to correct the app, as they would only need to access the functionality when they verified an answer misinterpreted by MathAIde.

In this sense, teachers printed the list of exercises before each class. For reasons of space and the goal of this article, we will only focus on the corrections made by teachers through the new functionality implemented and not on the exercise in each list. However, to be clear, each list applied by each teacher was recommended by MathAIde and contained four math exercises based on BNCC skills:

- **List 1 (L1) and List 2 (L2):** Skill EF02MA06 – Solve and elaborate addition and subtraction problems involving numbers of up to three orders, with the meanings of joining, adding, separating, removing, using personal or conventional strategies.
- **List 3 (L3) and List 4 (L4):** Skill EF03MA07: Solve and elaborate multiplication problems (by 2, 3, 4, 5, and 10) with the meanings of adding equal portions and elements presented in a rectangular arrangement, using different calculation strategies and records

Then, all captures and records made by the study teachers were recorded in the application database, allowing subsequent analysis. In total, 784 student answers were collected. Additionally, we conducted a post-study interview with the three teachers to capture insights into using MathAIde's AuI functionality.

### 7.3. Data analysis and results

#### 7.3.1. Student answers and usage of AuI

Using MathAIde's database, we analyzed the application's answers for each student's solution. The rationale is that the system's AI has known limitations, as reported in previous research Rodrigues, Guerino, et al. (2024); Chevtchenko et al. (2024), which guided us to focus this study on users' perceptions. Therefore, we analyzed how often the teacher needed to use the AuI functionality proposed in this work. Table 3 shows the results obtained from this analysis. Answers marked with "x" mention exercises that students got wrong, while those marked with "✓" mention those that they got right. The SR (student got it right) column shows how often the teacher used AuI to say that the student got an answer right and that MathAIde had told them they got it wrong. The CE (change error) column shows how often the teacher used AuI to change the student's error and that MathAIde had mentioned another.

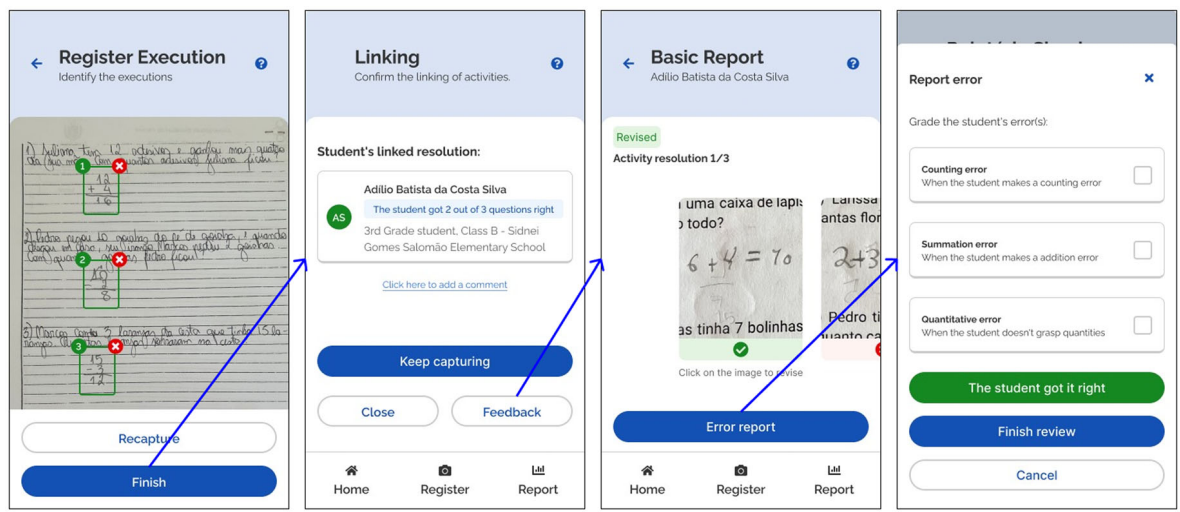

**Figure 4.** Flow to be followed to correct an answer. On the "record execution" screen, the teacher takes a photo of the student's notebook and links the answers to the exercises; on the "linking" screen, the teacher receives information from the linked student; when clicking on "feedback" button, the teacher is redirected to the "basic report" screen, which shows each student's answer separately, and whether the student got it right or wrong according to the AI; when clicking on "report error" button, a drawer opens where the teacher can correct the answer.

Through Table 3, we noticed that the functionality that allows AuI was used by teachers in the case study. Regarding the correction indicating that the student got it right (column SR in Table 3), we observed that 130 answers were corrected through this functionality, representing 16.5% of the total answers captured. Regarding the correction that changes the error made by the student (column CE in Table 3), we note that nine answers had their errors changed, representing 1.1% of the total answers

Table 3. Response to students' answer and frequency of use of the AuI functionality.

| Student | L1 | | | | L2 | | | | L3 | | | | L4 | | | | AuI | |
|---|---|---|---|---|---|---|---|---|---|---|---|---|---|---|---|---|---|---|
| | A1 | A2 | A3 | A4 | A5 | A6 | A7 | A8 | A9 | A10 | A11 | A12 | A13 | A14 | A15 | A16 | SR | CE |
| S01 | x | x | ✓ | ✓ | x | x | x | x | x | x | x | ✓ | x | x | x | x | 0 | 0 |
| S02 | x | x | ✓ | ✓ | x | x | x | x | x | ✓ | ✓ | ✓ | x | x | x | x | 0 | 0 |
| S03 | ✓ | ✓ | ✓ | x | x | x | x | x | ✓ | ✓ | ✓ | x | x | x | x | x | 6 | 0 |
| S04 | x | x | x | x | x | x | x | x | x | x | ✓ | ✓ | x | x | x | x | 0 | 0 |
| S05 | ✓ | ✓ | ✓ | ✓ | ✓ | ✓ | ✓ | ✓ | ✓ | ✓ | ✓ | ✓ | ✓ | ✓ | ✓ | ✓ | 0 | 0 |
| S06 | x | x | x | x | x | x | x | x | x | x | ✓ | x | x | x | x | x | 0 | 0 |
| S07 | ✓ | ✓ | ✓ | x | x | x | x | x | ✓ | ✓ | ✓ | x | x | x | x | x | 6 | 0 |
| S08 | x | x | x | x | x | x | x | x | ✓ | x | x | x | x | x | x | x | 0 | 0 |
| S09 | x | x | x | x | x | x | x | x | | x | x | x | x | x | x | x | 0 | 0 |
| S10 | x | x | x | x | ✓ | ✓ | ✓ | ✓ | x | x | x | x | x | x | x | x | 4 | 0 |
| S11 | ✓ | ✓ | ✓ | ✓ | ✓ | ✓ | ✓ | ✓ | x | x | x | x | x | x | x | x | 4 | 0 |
| S12 | x | x | x | x | x | x | x | x | ✓ | ✓ | x | x | x | x | x | x | 0 | 0 |
| S13 | ✓ | ✓ | ✓ | ✓ | x | x | x | x | x | x | x | x | x | x | x | x | 0 | 0 |
| S14 | ✓ | ✓ | ✓ | ✓ | ✓ | ✓ | ✓ | ✓ | ✓ | ✓ | ✓ | ✓ | ✓ | ✓ | ✓ | ✓ | 16 | 0 |
| S15 | x | x | x | x | x | x | x | x | x | x | ✓ | ✓ | x | x | x | x | 0 | 0 |
| S16 | x | x | x | x | ✓ | ✓ | ✓ | ✓ | x | x | x | x | x | x | x | x | 4 | 0 |
| S17 | x | x | x | x | x | x | x | x | x | ✓ | ✓ | ✓ | x | x | x | x | 0 | 0 |
| S18 | x | x | x | x | x | x | x | x | x | x | x | x | x | x | x | x | 0 | 0 |
| S19 | x | x | x | x | x | x | x | x | x | x | x | x | x | x | x | x | 0 | 0 |
| S20 | x | x | x | x | x | x | x | x | ✓ | x | x | x | x | x | x | x | 0 | 0 |
| S21 | ✓ | ✓ | ✓ | ✓ | ✓ | ✓ | ✓ | ✓ | ✓ | ✓ | ✓ | ✓ | ✓ | ✓ | ✓ | ✓ | 0 | 0 |
| S22 | x | x | x | x | x | x | x | x | x | ✓ | ✓ | ✓ | x | x | x | x | 0 | 0 |
| S23 | x | x | x | x | x | x | x | x | x | x | x | ✓ | x | x | x | x | 0 | 0 |
| S24 | x | x | x | x | ✓ | ✓ | ✓ | x | x | x | x | x | x | x | x | x | 3 | 0 |
| S25 | x | x | x | x | x | x | x | x | x | x | x | ✓ | x | x | x | x | 0 | 0 |
| S26 | x | x | x | x | x | x | x | x | x | x | x | ✓ | x | x | x | x | 0 | 0 |
| S27 | x | x | x | x | x | x | x | x | x | ✓ | x | ✓ | x | x | x | x | 0 | 0 |
| S28 | ✓ | ✓ | ✓ | ✓ | ✓ | ✓ | ✓ | ✓ | ✓ | ✓ | ✓ | ✓ | ✓ | ✓ | ✓ | ✓ | 16 | 0 |
| S29 | ✓ | ✓ | ✓ | ✓ | ✓ | ✓ | x | x | x | x | x | x | x | x | x | x | 3 | 0 |
| S30 | ✓ | ✓ | ✓ | ✓ | ✓ | ✓ | ✓ | ✓ | ✓ | ✓ | ✓ | ✓ | ✓ | ✓ | ✓ | ✓ | 0 | 0 |
| S31 | x | x | x | x | x | x | x | x | x | x | ✓ | ✓ | x | x | x | x | 0 | 0 |
| S32 | ✓ | ✓ | ✓ | ✓ | ✓ | ✓ | ✓ | ✓ | ✓ | ✓ | ✓ | ✓ | ✓ | ✓ | ✓ | ✓ | 0 | 0 |
| S33 | ✓ | ✓ | ✓ | ✓ | ✓ | ✓ | ✓ | ✓ | ✓ | ✓ | ✓ | ✓ | ✓ | ✓ | ✓ | ✓ | 0 | 0 |
| S34 | x | x | x | x | x | x | x | x | x | x | x | x | x | x | x | x | 0 | 0 |
| S35 | ✓ | ✓ | ✓ | ✓ | ✓ | ✓ | ✓ | ✓ | ✓ | ✓ | ✓ | ✓ | ✓ | ✓ | ✓ | x | 12 | 1 |
| S36 | ✓ | ✓ | ✓ | x | x | x | x | x | ✓ | ✓ | ✓ | x | x | x | x | x | 6 | 0 |
| S37 | ✓ | ✓ | ✓ | x | x | x | x | x | ✓ | ✓ | ✓ | x | x | x | x | x | 6 | 0 |
| S38 | ✓ | ✓ | ✓ | x | ✓ | ✓ | x | x | ✓ | ✓ | ✓ | x | ✓ | ✓ | ✓ | x | 8 | 4 |
| S39 | ✓ | ✓ | ✓ | ✓ | ✓ | ✓ | ✓ | ✓ | ✓ | ✓ | ✓ | ✓ | ✓ | ✓ | ✓ | ✓ | 16 | 0 |
| S40 | ✓ | ✓ | ✓ | ✓ | ✓ | ✓ | ✓ | ✓ | ✓ | ✓ | ✓ | ✓ | ✓ | ✓ | ✓ | ✓ | 0 | 0 |
| S41 | ✓ | ✓ | ✓ | ✓ | ✓ | ✓ | ✓ | ✓ | ✓ | ✓ | ✓ | ✓ | ✓ | ✓ | ✓ | ✓ | 0 | 0 |
| S42 | ✓ | ✓ | ✓ | ✓ | ✓ | ✓ | ✓ | ✓ | ✓ | ✓ | ✓ | ✓ | ✓ | ✓ | ✓ | x | 12 | 1 |
| S43 | ✓ | ✓ | ✓ | x | x | x | x | x | x | x | x | x | x | x | x | x | 0 | 1 |
| S44 | ✓ | ✓ | ✓ | ✓ | ✓ | ✓ | ✓ | x | ✓ | ✓ | ✓ | ✓ | ✓ | ✓ | ✓ | x | 8 | 2 |
| S45 | x | x | x | x | x | x | x | x | x | ✓ | ✓ | ✓ | x | x | x | x | 0 | 0 |
| S46 | ✓ | ✓ | ✓ | ✓ | ✓ | ✓ | ✓ | ✓ | ✓ | ✓ | ✓ | ✓ | ✓ | ✓ | ✓ | ✓ | 0 | 0 |
| S47 | x | x | x | x | x | x | x | x | x | x | x | x | x | x | x | x | 0 | 0 |
| S48 | ✓ | ✓ | ✓ | ✓ | ✓ | ✓ | ✓ | ✓ | ✓ | ✓ | ✓ | ✓ | ✓ | ✓ | ✓ | ✓ | 0 | 0 |
| S49 | ✓ | ✓ | ✓ | ✓ | ✓ | ✓ | ✓ | ✓ | ✓ | ✓ | ✓ | ✓ | ✓ | ✓ | ✓ | ✓ | 0 | 0 |

Acronyms – L: lists; A: activities; AuI: Augmented Intelligence; SR: student got it right; CE: change error. The SR column shows how often the teacher used AuI to mention that the student got an answer right and that MathAIde had told her (him) she (he) got it wrong. The CE column shows how often the teacher used AuI to change the student's error and that MathAIde had mentioned another.

obtained. In this sense, the functionality allowing AuI was used 139 times in total, representing 17.6% of the total answers. These findings suggest the usefulness of the functionality, though further validation is required with larger and more diverse samples to confirm its generalizability.

For example, student S14 had his answers corrected 100% of the time. In other words, in all their answers, the teacher needed to indicate that they got the solution right. The fact that MathAIde got this student's answer wrong may be linked to several factors involving the context of the application: classroom lighting, quality of the photo taken by the teacher, student writing, and training of the AI model, among others. The fact is that allowing the teacher to work together with the AI benefited the student and made the assessment carried out by the application fairer. In this sense, the results of this study point to the high use of the functionality that allows AuI in MathAIde and highlights its importance for a fairer and more assertive ITS, helping and working with the teacher.

### 7.3.2. Post-study interviews with teachers

We performed a thematic analysis of the interview transcripts. As teachers already knew the functionality and its positive points from participating in the A/B Test, we focused the analysis of this study on understanding the correction behavior in a real environment. From the analysis, we identified two main themes: (i) feedback and correction of the MathAIde and (ii) improvements in the correction functionality.

Participants expressed a preference for a more objective and faster feedback system. They suggested that AI should focus on evaluating and providing feedback more assertively, avoiding excessive classification of errors that do not contribute to learning assessment (see citation C11). Participants also proposed a color classification to indicate whether the answer was correct (green), incorrect (orange), or if there was something extra to note (red) (see citation C12).

- C11: "There were four options (to report an error), and you had to read them. You just registered for the app, and the student quickly got it right. This error indication is far beyond the proposal. It would be interesting to be more objective and correct if a student got it right or wrong. I need the app to help me evaluate it." (P1)
- C12: "When we take the photo, it could be color-classified to understand the student's classification. For example, if s/he got it right, the square turns green. If s/he makes a mistake, it turns red. Because then we would know which answer we can correct, saving time." (P2)

Participants mentioned that they trusted the app's feedback system, assuming it was correct because the options were marked as correct (green) previously. They suggested that the color of the options marked as correct influenced their confidence in the feedback system, leading them to believe there was no error even without checking in detail (see citation C13).

- C13: "As is the case at school, I did it so quickly that in some cases, I did not press the option to see the feedback or report some errors. As some were green, I trusted that the student had gotten it right." (P2)

Participants also highlighted the importance of language that points directly to the review, suggesting a word that semantically indicates this action (see citation C14). Furthermore, participants emphasized simplifying the correction process to avoid work overload, especially in classrooms with many students. They proposed implementing improvements that allow cutting the flow of the correction process, making it more efficient and less laborious (see citation C15).

- C14: "I do not think the flow is wrong. It is correct. I think the only thing that could improve is a word that semantically points directly to review because "report error" continues to look like an error on the part of the app manufacturer and not in identifying the answer." (P1)
- C15: "The application could show the error by color and allow us to report it right when taking the photo as soon as we see it. There could be this option to classify by color and allow correction." (P2)

### 7.4. Synthesis

This study showed that the functionality that allows AuI in MathAIde through system correction was used in the real environment, which allowed a fairer application and greater freedom for the teacher in the STI. However, the interview revealed that there is still a long way to go, and improvements can be implemented to improve functionality.

## 8. Discussion

Our RQ asked how to design AuI so that it empowers mobile ITS supporting handwritten input. In answering it, this article detailed a user-centered mixed methods approach that led to a usable, user-validated AuI feature that was successfully implemented in MathAIde, a mobile ITS supporting handwritten input that helps in mathematics education. To unpack how our results lead to this finding, this section interprets the findings from the four studies, along with their implications and contributions, then discusses threats to validity and how we mitigated them.

### 8.1. Findings interpretation

By applying a user-centered approach to propose AuI in an ITS, MathAIde in our case, we found that the mixed-methods approach enabled us to: (i) offer several alternatives based on the vision of potential end-users; (ii) foster a culture of prioritization and balance between technological feasibility and user needs; (iii) enable the use of specific metrics to validate a proposal based on the context in which the user is inserted; and (iv) verify the usefulness and use of the solution.

Regarding the different solution possibilities, our results showed some ways to enable AuI in the context of MathAIde. Based on the vision of potential end users, we realized that the solution could combine several components, such as colors, shapes, and buttons Barros and Melo (2011). Furthermore, users also provided possibilities that involve a high level of control and a high level of automation, such as editing misidentified numbers or changing errors made by students, which are characterized in quadrant Q4 of the framework proposed by Shneiderman (2022).

Despite the proposed options, we had to prioritize and validate the proposals through prototyping and A/B testing. This prioritization and validation were essential to balance user needs with technological feasibility. Despite the different solutions proposed by the brainstorming participants, the solution to be implemented should be something that could be developed considering the technological capacity of the application. In this sense, it was necessary to understand the technological capabilities to adjust the identified solutions and verify what was possible and what was not possible to be done. So, this process was essential to align expectations and capabilities in the development process.

Regarding the specific metric, the mixed user-centered approach showed us that the time to use the solution was a determining factor in the proposal's success. Our results showed that the solution that provided pre-defined correction alternatives for teachers proved to be the most efficient and preferred in the A/B test. Although the solution that involved correcting the numbers provided control for the user, it took more time to carry out, which could be a problem in large classes and low technological knowledge. The results then showed that enabling correction and providing pre-defined help was the balance between enabling AuI in MathAIde and the ITS's efficiency.

In this sense, the study in a real environment was essential to verify that the proposal was helpful for teachers, considering the number of times it was used to indicate that a student got it right and to change the error made. The best scenario is if MathAIde does not make mistakes. However, we recognize the technological limitations and variations in students' answers, as they often use drawings and scribbles to answer questions, making recognition by the application difficult. In this way, the solution helped reduce the gap involved in the correction automation process.

### 8.2. Implications

This article has several implications. We demonstrate how a user-centered, mixed-methods approach Guimarães et al. (2017) contributes to designing AuI-based ITS. While past research has similarly explored high levels of both user control and computational automation, to our best knowledge, previous studies have not addressed these design challenges in the context of AIED Alfredo et al. (2024); Topali et al. (2025); Lawrence et al. (2024). Differently, we included teachers in all design phases, which enabled us to find, prioritize, design, and evaluate several insights aligned with our target users' contexts and needs. Our article contributes insights into how including target users in all design phases and complementary methods improved the final design.

Furthermore, we demonstrate that our design approach led to usable design. Although the literature argues on the importance of including users in all design phases, many AIED studies do not follow this approach Alfredo et al. (2024) or adapt it based on pre-defined models that might fail to capture nuances of users and contexts' complexity Shreiner and Guzdial (2022). This might lead to unpackable or adopted products, preventing the target users from benefiting from AIED systems as expected Chiu et al. (2023). In contrast, this article presents a teacher-centered design solution for ITS based on AuI, which is prominent in improving and empowering teachers with a high level of control while benefiting from computational automation provided by ITS. Our article contributes a usable, teacher-centered design that might be reused and expanded upon.

Implementing a user-centered approach in an existing tool, MathAIde, corroborates the growing importance of aligning AI development with the practical realities of educators and learners Chiu et al. (2023). The system was tailored to authentic classroom needs by involving potential end-users rather than abstract technological possibilities. This reaffirms a core principle in AIED: effective educational AI must be pedagogically aligned and socially embedded Pedro et al. (2019); Vinuesa et al. (2020). The ability to generate design alternatives grounded in teachers' perspectives demonstrates how the user-centered approach fosters innovation and relevance, positioning it as a prominent path for future ITS and AuI developments.

Accordingly, one of the study's key contributions lies in its structured process for managing the trade-off between user needs and technological feasibility. Despite various user-generated solutions, only those that could be implemented within the system's existing capabilities were advanced. This reflects a mature approach to AIED development, where constraints are not seen as barriers but as drivers of creativity and prioritization. The iterative validation process, including prototyping and A/B testing, offers a replicable model for other ITS designers, emphasizing that educational impact emerges from innovation and implementable innovation that bridges aspirational design and operational realities.

Moreover, this study highlights design flexibility's relevance. Findings related to user preferences in the tool's design reiterate the value of flexible interfaces in educational technologies. Additionally, the dual need, including high control and automation, reflects a sophisticated user requirement for situational adaptability. These findings echo Shneiderman (2022)'s framework on human-centered AI and call for ITSs that are not rigidly designed but capable of dynamically responding to user expertise, task complexity, and learning context. This flexibility will be essential for scaling AIED tools across diverse educational settings with varying levels of digital literacy.

Lastly, recognizing current limitations in AI capabilities, especially in handling diverse student input forms like drawings and scribbles, demonstrates a reasonable realism in system evaluation. This acknowledgment is crucial in a field often driven by idealized performance benchmarks Chiu et al. (2023). It reinforces the need for hybrid intelligence models like AuI, where AI augments but does not replace human oversight, particularly in contexts where interpretive ambiguity is high Sadiku et al. (2021); Yau et al. (2021). This finding calls for further research into robust, context-aware input recognition. It highlights the value of fallback mechanisms that maintain system utility even in the face of imperfect automation.

### 8.3. Threats to validity

*Internal validity* threats primarily revolve around potential history and maturation effects. During the Brainstorming phase, the evolving nature of discussions and participants' familiarity with the problem might introduce historical biases, impacting the ideation outcomes. Moreover, participants' cognitive development or changing perspectives throughout the study could affect the quality and novelty of proposed solutions. To counter these threats, we sought diverse participant backgrounds and employed structured facilitation to ensure that ideas were evaluated without being influenced by external factors. Concerning *external validity*, the risk of selection bias is prominent in our research due to the specific nature of participants involved in each study phase. For instance, the Prototyping phase involves a smaller subset of individuals who may not fully represent the broader user base or educational contexts where MathAIde is utilized. This limitation could affect the generalizability of our findings and the transferability of proposed solutions to different educational settings. To mitigate this threat, we thoroughly assessed users' needs and engaged participants with varied educational backgrounds to capture a more comprehensive range of perspectives.

Concerning *construct validity*, ensuring the accuracy and consistency of measurement instruments and constructs is crucial in our article's context. The Prototyping phase, where functional requirements are translated into interface prototypes, is particularly susceptible to measurement biases if the identified requirements do not align accurately with user expectations or operational needs. Additionally, construct confounding might arise if the proposed solutions lack clarity or if multiple variables influence user interactions with MathAIde. To address these concerns, we employed iterative refinement processes, solicited feedback from domain experts, and utilized established usability testing frameworks to validate the functionality and usability of prototypes. Concerning *conclusion validity*, the Case Study phase presents challenges related to conclusion validity, particularly regarding ambiguous temporal precedence and selection-maturation interactions. As teachers interact with MathAIde in real-world classroom settings over several days, external events or pedagogical changes could impact their perceptions and corrective behaviors. Moreover, distinguishing the effects of prolonged exposure versus inherent improvements in MathAIde's performance poses a methodological challenge in drawing accurate conclusions about the efficacy of the correction functionality. To address these issues, we implemented pre- and post-study interviews, collected interaction logs, and collaborated closely with educators to contextualize observed outcomes within broader teaching practices and student learning dynamics.

## 9. Conclusion

This study addressed the gaps in designing and implementing AIED systems by studying the MathAIde ITS. While ITSs are one of the most effective kinds of AIED systems, we chose MathAIde as a proxy to study ITSs that do not require students to have access to technological devices, as those provide an alternative interface that helps expanding the accessibility of such tools to low-resource contexts. Hence, by incorporating a user-centered mixed-methods approach, we have demonstrated how AuI can be integrated into ITSs.

One of the main limitations of this study is the small and localized sample size, particularly in the A/B Test and Case Study phases. Although participants brought valuable insights, their demographic and pedagogical contexts may not fully represent the diverse realities of elementary school mathematics teachers. Future work should explore the generalizability of our findings in broader and more varied educational contexts.

Our research has several key takeaways: (i) User-centered design is essential: by involving teachers in all design phases, we ensured that the resulting ITS system was practical and aligned with the users' needs. The collaborative process allowed us to explore multiple design possibilities and prioritize those that balanced user requirements with technological feasibility; (ii) Mixed-methods approach yields practical solutions: combining brainstorming sessions, high-fidelity prototyping, A/B testing, and real-world classroom case studies provided comprehensive insights into user preferences and system performance. This iterative process helped refine the ITS system, making it more responsive to the real-world educational environment; (iii) Balancing control and automation: our findings underscored the importance of balancing user control with computational automation. The preferred solution provided teachers with predefined correction alternatives, which enhanced efficiency without compromising the level of control needed for accurate assessment. This balance is crucial for adopting and sustaining AIED systems in diverse educational settings; (iv) Real-world validation: implementing and testing the solution in classroom environments validated its practicality and usefulness. Teachers could use the system to correct and provide feedback on students' work despite the inherent challenges of varied student inputs. This practical application confirms the potential of ITS systems to support education in resource-constrained environments.

In future work, we intend to improve MathAIde's AUI further based on the insights from the case study in a real environment. Additionally, we will conduct large-scale studies with MathAIde to check other metrics such as effectiveness, efficiency, learning analytics, and learner and teacher experience. While initial findings indicate usability and relevance, long-term adoption and scalability remain open challenges. Ensuring continued teacher engagement will require further integration with school curricula, broader AI error-handling capabilities, and localized support. Additionally, we plan to explore integration with learning analytics dashboards and teacher training materials to enhance sustained use.

## Notes

1. In the context of this paper, we interpret human-centered and user-centered interchangeably, given that AIED system users are humans, to respect Shneiderman (2022)'s terminology in proposing their framework.
2. https://zenodo.org/records/11106385.
3. https://zenodo.org/records/11106426.

## Acknowledgements

We wish to express our gratitude to everyone involved in this national project, including researchers, policy-makers, students, and teachers.

## Author contributions

CRediT: **Guilherme Guerino**: Conceptualization, Data curation, Formal analysis, Methodology, Writing – original draft, Writing – review & editing; **Luiz Rodrigues**: Data curation, Formal analysis, Methodology, Writing – original draft, Writing – review & editing; **Luana Bianchini**: Investigation, Visualization; **Mariana Alves**: Investigation, Validation; **Marcelo Marinho**: Project administration, Supervision, Writing – review & editing; **Thomaz Veloso**: Methodology, Resources, Writing – review & editing; **Valmir Macario**: Project administration, Supervision, Writing – review & editing; **Diego Dermeval**: Funding acquisition, Resources, Writing – review & editing; **Thales Vieira**: Funding acquisition, Resources, Writing – review & editing; **Ig Bittencourt**: Funding acquisition, Resources, Writing – review & editing; **Seiji Isotani**: Funding acquisition, Resources, Writing – review & editing.

## Disclosure statement

The authors report there are no competing interests to declare.

## Funding

This work was supported by the Brazilian Ministry of Education (MEC) – TED 11476.

## ORCID

Guilherme Guerino http://orcid.org/0000-0002-4979-5831
Luiz Rodrigues http://orcid.org/0000-0003-0343-3701
Luana Bianchini http://orcid.org/0009-0000-2185-2905
Mariana Alves http://orcid.org/0009-0003-3493-5664
Marcelo Marinho http://orcid.org/0000-0001-9575-8161
Thomaz Veloso http://orcid.org/0000-0003-0889-7564
Valmir Macario http://orcid.org/0000-0002-7816-5759
Diego Dermeval http://orcid.org/0000-0002-8415-6955
Thales Vieira http://orcid.org/0000-0001-7775-5258
Ig Bittencourt http://orcid.org/0000-0001-5676-2280
Seiji Isotani http://orcid.org/0000-0003-1574-0784

## About the authors

**Guilherme Guerino** is a professor at the State University of Paraná, and a Researcher at the State University of Maringá, Brazil. He serves as Head of Innovation and Technology in his institution and conducts research in the field of HCI, focusing on UX methods and techniques, innovation, and UCD.

**Luiz Rodrigues** is a professor at the Federal University of Technology of Paraná, Brazil. His interests are focused on the design, development, and experimentation of educational technologies, utilizing artificial intelligence techniques and prioritizing user experience. His main research areas include Artificial Intelligence in Education, Gamified Learning, and Learning Analytics.

**Luana Bianchini** holds a Master's degree in Computer Science from the Federal University of São Carlos, Brazil, where she conducted research in UX. She has experience in UX Research, conducting remote and in-person studies. She currently works in UX Design and Research, performing research during the software product development process.

**Mariana Alves** is a UX Designer and a master's student in Applied Computing. She is a researcher in HCI and Augmentative and Alternative Communication (AAC). She has experience in projects that apply Artificial Intelligence to Education, focusing on accessibility, usability, and the development of user-centered technologies.

**Marcelo Marinho** is a professor at the Federal Rural University of Pernambuco, Brazil. He works in Software Engineering and Information Systems, developing research and projects primarily in the following areas: software project management, global software development, agile methods, innovation in software projects, and the software development process.

**Thomaz Veloso** holds a PhD in Teleinformatics Engineering and Science from the Federal University of Ceará, Brazil, and the University of Copenhagen, Denmark, respectively. He has experience in educational assessment, teacher training, data mining, pattern recognition, multivariate statistics, and linear and multilinear algebra applied to problems in Educometrics and Engineering.

**Valmir Macario** is a researcher at the Federal University of Pernambuco and a professor at the Federal Rural University of Pernambuco, Brazil. He has experience in Computer Science, with an emphasis on Intelligent Computing (Artificial Intelligence), working primarily on the following topics: semi-supervised learning and image processing.

**Diego Dermeval** is a professor at the Federal University of Alagoas, a member of the Advisory Board of the Center for Excellence in Social Technologies, and a visiting scholar at the Harvard Graduate School of Education (USA). His research focuses on AI in Education and Intelligent Tutoring Systems, focusing on supporting teachers and students in resource-constrained settings.

**Thales Vieira** is a professor at the Federal University of Alagoas, Brazil. He has experience in Mathematics and Computer Science, with an emphasis on Visual and Intelligent Computing. He has researched solutions to complex problems in Computer Graphics and Computer Vision, as well as Natural Language Processing (NLP).

**Ig Bittencourt** is a professor at the Federal University of Alagoas, Brazil, and a visiting scholar at the Harvard Graduate School of Education. He has dedicated his efforts to education, focusing on the design, development, and experimentation of intelligent educational technologies to promote equitable learning and well-being.

**Seiji Isotani** is a professor at the University of São Paulo and the Berkman Klein Center at Harvard University, as well as a visiting scholar at the Graduate School of Education, University of Pennsylvania. He is a leading figure in the fields of gamification, artificial intelligence in education, and educational technologies.